# Continuously tuning the refractive indices by restructuring materials on the 20-1000 atoms scale: improving anti-reflection coating designs

**Jacob Poole[1,*], Aidong Yan[1], Paul Ohodnicki[2], and Kevin Chen[1]**

[1]*Department of Electrical and Computer Engineering, University of Pittsburgh Pittsburgh, PA, 15261, USA*
[2]*National Energy Technology Laboratory, 626 Cochran Mill Road, Pittsburgh, PA 15236, USA*
[*]*JakePooleFin@gmail.com*

**Abstract:** We demonstrate the capability of block-copolymer templating to tune the refractive indices of functional oxides over a broad range by structuring materials on the 20-1000 atoms scale, with simple one-pot synthesis. The presented method is then combined with genetic algorithm-based optimization to explore its application for anti-reflection coating design. Merging these techniques allows for the realization of a minimal two-layer anti-reflection stack for silicon with broadband reflectivity of just 3% from the nominal value of 38%, over a 120° angular span, validated by fabrication followed by optical measurements.

## 1. Introduction

Nature provides a sparse distribution of functional materials with useful refractive indices. It is often the case that designs must be formulated around existing material properties which are highly limited in range. In optics and photonics, the refractive index is specifically chosen to obtain a desired functionality as it defines the working relationship between light and matter. Existing methods, such as doping, give access to some control over the choice of the available refractive indices. However, the range of control afforded by these methods is quite small, a few percent at best. Magnesium fluoride (n~1.37) is the only functional non-lossy material between glass (n~1.45) and air (n~1). This sparsity in the refractive indices severely limits our capabilities in optical design and exploration.

There are several areas that could benefit from having a greater refractive index availability, especially if the manufacturing technique is simple, such as solution processing. This capability could open many research avenues by providing easier methods for the manufacture of structures whose refractive indices can substantially vary in 1D, 2D, and 3D. A thin-film stack would be a 1D refractive index structure, whereas embedded cylinders and spheres constitute the 2D and 3D variations.

From theory, an optimal anti-reflection coating would consist of a continuously varying refractive index from the substrate to the background (1D refractive index gradient)[1, 2]. For silicon, this would require having materials with refractive indices continuously distributed between 3.5 and 1, to truly minimize light reflections. A number of methods have been demonstrated such as nano-arrays[3-5], surface textures[6-8], anodic alumina[9], lithography and wet etching[10], and a combination of oblique angle deposition with sputtering techniques[1, 11, 12] to address the need of improving anti-reflection technology. Many of these are based on technologies such as e-beam, CVD, and sputtering that do not scale well and are highly restrictive in deposition. Currently, the most used anti-reflection coatings are still single layer $Si_3N_4$ and $TiO_2$, which reduce reflections to ~18% over the solar conversion window[11]. Having a continuous availability in the refractive indices would give access to control over the angular span in the optimization process and greater control over the wavelength span, in addition to just optimization at normal incidence.

Transformation optics and quasi conformal-mapping are an upcoming and interesting design methodology in optics and photonics[13]. With these techniques a variety of designs have been carried out to allow for the unconventional manipulation of light for practical applications, such as in light wave circuits and silicon photonics[14]. It is a tensor based design method[15, 16], which, in most cases, provides designs that require strong 2D and 3D refractive index gradients. These are difficult if not impossible to realize with conventional techniques at optical frequencies. Therefore, a solution processing route that provides extensive control over the refractive indices could be used to address the manufacturability issues of 2D and 3D refractive index gradients.

In this paper the refractive indices of thin-films of $TiO_2$, $ZnO$, $SnO_2$, and $SiO_2$ are varied continuously in the range of 1.17 to 2.2. The method employed combines block-copolymer based templating with the highly scalable and flexible solution processing route. To demonstrate the potential of the presented technique, a broadband and omnidirectional anti-reflection stack was designed using genetic algorithm based optimization. With just a two-layer design the reflectivity of silicon was reduced from its nominal value of ~38% down to 3%, over a 120° angular span. Reducing the angular dependence with antireflection coatings could eliminate the need for sun tracking in solar-cells as most sunlight falls within this angular range.

*1.1 Tailoring the Refractive Indices*

The idea to manipulate materials to tailor their dielectric permittivity dates back to the Clausius-Mosotti relation(1850)[17, 18] and later to the Maxwell-Garnet (1904)[19] and the Bruggeman (1935)[20] effective medium theories. In the quasi-static regime, where sufficient separation exists between the wavelength of light and the geometric features of the compositions (~$\lambda/10$), these concepts can be realized by creating 3D material mixtures. The effective medium theories attempt to establish a link between a microscopic quantity (atomic polarizability) and a macroscopic quantity (dielectric constant) when material mixtures alternate with spacing much less than the wavelength of light (<20nm)[21]. The realization of these theories provide access to engineering the optical properties of materials at the near-atomic scale.

Block-copolymer templating is a well explored method to structure materials for a variety of applications in Chemistry, with features in the 5-100nm range[22-24]. For exploring the 3D structuring of functional materials in the deep sub-wavelength regime (<20nm) a non-ionic hydrophilic triblock-copolymer, Pluronic F-127, is chosen. This triblock-copolymer is known to have a higher temperature stability and to form 3D structures with the metal alkoxide wet processing route[24-27]. After the one-pot synthesis of a coating solution and deposition, the block-copolymer metal-alkoxide composite is transformed into a functional material matrix with structure features on the 5-20nm scale. Varying the molecular ratio between the block-copolymer and the metal alkoxide provides variations in size[28] and distribution of the formed structures, giving way to controllable refractive indices.

*1.2 Anti-Reflection Coating Design*

Quarter-wave thin film stacks are very efficient at minimizing reflections but at the high cost of greatly reduced angular and wavelength spans. Optimal anti-reflection coatings predicted by theory that minimize light reflections globally over a large angular and wavelength span require materials with continuous refractive index profiles from the substrate to the background refractive index (1D refractive index gradients). Refractive index profiles that minimize reflections have been previously explored analytically[29] and by numerical optimization methods, such as the genetic algorithm[1, 30, 31], and the simulated annealing algorithm[32]. Continuous profiles are impossible to realize therefore practical approximations to these are multi-layered coatings consisting of a few layers, designed by the optimization coupled transfer matrix method[1, 31, 33].

The transfer matrix method was used to determine the amount of the reflected light as a function of wavelength, angles of incidence, polarization state, and the thickness and refractive index of each layer. Bruggeman's effective medium theory (**Equation 1**) was used to link the volume fraction of the nanomaterial air 3D mixture with the refractive indices[21]. Where, f is the volume fraction of air, $\varepsilon_1$ and $\varepsilon_2$ are the permittivity of the host and the inclusions, which in this case is metal oxide and air, $\varepsilon_{eff}$ is the resultant permittivity, and the refractive index n= $\sqrt{\varepsilon_{eff}}$.

$$f\left(\frac{\varepsilon_1 - \varepsilon_{eff}}{\varepsilon_1 + 2\varepsilon_{eff}}\right) + (1-f)\left(\frac{\varepsilon_2 - \varepsilon_{eff}}{\varepsilon_2 + 2\varepsilon_{eff}}\right) = 0 \qquad \text{Equation 1.}$$

The fitness function used for evaluating the quality of the found solution is given by **Equation 2**. For reducing the reflectivity, the parameter to be optimized is the average reflectance over wavelength and angles of incidence[31]. Here $R_{TE}$ and $R_{TM}$ are the two polarization components of the Fresnel equations. As previously suggested, a weight function $w(\lambda,\theta)$ can be used to preferentially adjust the parameters to account for, for example, variations in light intensity in the solar spectrum as a function of wavelength and angles of incidence or for variations in the internal/external quantum efficiency of solar cells[31]. As a second measure the found solutions by the genetic algorithm were verified with the simulated annealing algorithm to ensure consistency.

$$R_{avg} = \frac{1}{\Delta\lambda\Delta\theta} \int_{\lambda_{min}}^{\lambda_{max}} \int_{\theta_{min}}^{\theta_{max}} \frac{1}{2} w(\lambda,\theta) \cdot (R_{TE} + R_{TM}) d\theta d\lambda \qquad \text{Equation 2.}$$

In the optimization the nominal refractive indices for silicon, titanium dioxide, and silicon dioxide were obtained from the literature[34-36]. Simulations were performed over two wavelength and angular ranges (**Table 1**): one in the visible spectrum (400-700nm) and over an angular span of 0-60, and a second over the solar conversion window (400-100nm) with and angular span of 0-75°. Horizon effects, which limit the angles at which sunlight can come to the earth, eliminate the need to consider angles over 75° from the normal.

**Table 1.** Optimization results for the thin-film based anti-reflection stack[a]

| (A) 400-700nm, and 0-60° | | | | (B) 400-1100nm, and 0-75° | | | | |
|---|---|---|---|---|---|---|---|---|
| Layers | 0 | 1 | 2 | 3 | Layers | 0 | 1 | 2 | 3 |
| V1 | | 83.5%TiO$_2$ | 51.1%SiO$_2$ | 40%SiO$_2$ | V1 | | 81.3%TiO$_2$ | 77.3%SiO$_2$ | 20%SiO$_2$ |
| V2 | | | 100%TiO$_2$ | 73.7%SiO$_2$ | V2 | | | 100%TiO$_2$ | 99%SiO$_2$ |
| V3 | | | | 100%TiO$_2$ | V3 | | | | 100%TiO$_2$ |
| H1 | | 63.94nm | 112.7nm | 200nm | H1 | | 88.45nm | 125.759nm | 223nm |
| H2 | | | 56nm | 120nm | H2 | | | 68.65nm | 107nm |
| H3 | | | | 57.4nm | H3 | | | | 68nm |
| Ravg | 38% | 6.64% | 2.80% | 3.60% | Ravg | 35% | 13.60% | 6.15% | 4.70% |

[a]The volume fractions V1-V3 of the respective materials and their film thicknesses H1-H3, optimized with the genetic algorithm, are listed from 0-3 layers. Optimizations were performed for two ranges, one for the visible spectrum (A) and another for the solar cell conversion window (B).

In the table $R_{avg}$ lists the average reflectance values obtain by optimization for the respective thin-film system. In the simulation for case A, given that the lowest currently attainable refractive index is 1.17, a third layer is detrimental in minimizing the reflectivity. The lowest reflectivity that can be obtained is 2.8% and that is with a two-layer system, given the refractive index bounds of ~2.4 ($TiO_2$) and 1.17 (Porous $SiO_2$). For case B, the addition of a third thin-film layer provides a considerable improvement. Plots of the simulation results for case A are given in **Figure 1**, where (A) shows the reflectivity of bare silicon as a function of wavelength and angle, and (B) shows the reflectivity over the same wavelength and angular span when an optimized two-layer thin-film coating is applied.

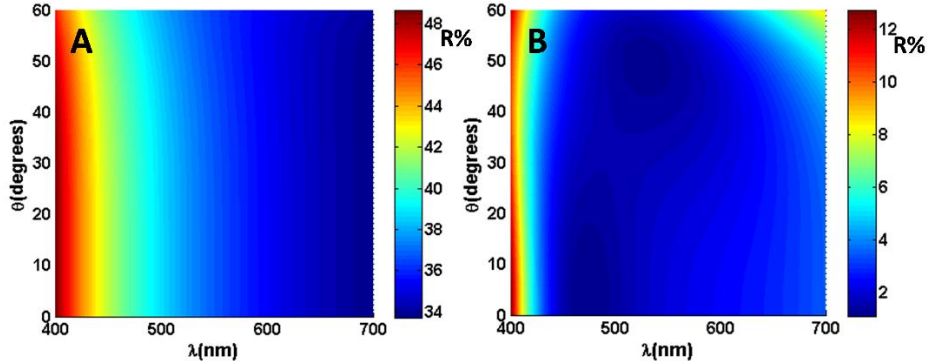

**Figure 1:** Light reflectivity of **A:** bare silicon and **B:** a thickness and refractive index optimized two-layer thin-film stack as a function of wavelength and angles of incidence.

### 2.0 Materials and Methods

Methods for the manufacture of the mesoporous $SnO_2$, $TiO_2$, ZnO, and $SiO_2$ nanomaterials using a non-ionic Pluronic type triblock-copolymer were adapted from the literature[26, 27, 37]. The manufacture of the ZnO nanostructure was obtained experimentally as the formation of ZnO nanostructures with the triblock-copolymer route is not well explored. First, a 32 mM Pluronic F-127 solution was prepared in 1-butanol to which 35 mmols of 37% HCl and 72 mmols of $H_2O$ were added. This solution was subsequently used as the polymer source. Various metal alkoxide solutions of $TiO_2$, $SiO_2$, $SnO_2$, and ZnO were prepared as per **Table 2**, which lists the constituents of the solutions in moles. The metal source for $TiO_2$ was titanium isopropoxide (TTIP), for $SiO_2$ the source was tetraethoxysilane (TEOS), for $SnO_2$ the source was $SnCl_4$, and for ZnO the source was zinc acetate dehydrate (Zn-acac). After preparation, the solutions were stirred at room temperature for 2 hours and left to age for one day before use. Dense $TiO_2$ was obtained from a solution of 0.7M titanium isopropoxide in 2-Methoxyethanol to which 2 moles of ethanolamine, with respect to the metal source, were added to stabilize the solution. The precursor solutions were deposited on ~2x2cm square pieces of <100> silicon wafer by the spin cast method at 2500 RPM. Followed by annealing at 400-800°C for 2 hours at a heating rate of 1°C/minute.

The optical properties of the samples were measured with a Jobin Yvon spectroscopic Ellipsometer. For the $SnO_2$, $SiO_2$ and ZnO nanostructures, the Lorentz oscillator model provided good fits to the measured data[38]. For the $TiO_2$ nanostructures the New Amorphous model was used to obtain the refractive indices[39]. Variations in the refractive indices were represented by changes in the porosity, modelled using Bruggeman's effective medium theory.

**Table 2.** Chemical composition of the thin-film polymeric precursor solutions

| TiO2 Solution | | | | | SiO2 Solution | | | |
|---|---|---|---|---|---|---|---|---|
| Sol# | TTIP | F-127 | HCl | Butanol | Sol# | TEOS | F-127 | HCl | Ethanol |
| Ti-1 | 1 | 0.004 | 0.54 | 28.4 | Si-1 | 1 | 0 | 2.86 | 45.22 |
| Ti-2 | 1 | 0.007 | 0.83 | 29.3 | Si-2 | 1 | 0.01 | 3.43 | 45.22 |
| Ti-3 | 1 | 0.01 | 1.12 | 30.4 | Si-3 | 1 | 0.01 | 4 | 49.74 |
| Ti-4 | 1 | 0.013 | 1.43 | 31.58 | Si-4 | 1 | 0.01 | 4.57 | 49.74 |
| Ti-5 | 1 | 0.016 | 1.43 | 32.47 | Si-5 | 1 | 0.01 | 6.8 | 51.2 |
| | | | | | Si-6 | 1 | 0.01 | 8 | 54.3 |
| | | | | | Si-7 | 1 | 0.01 | 10.28 | 63.3 |

| SnO2 Solution | | | | | ZnO Solution | | | |
|---|---|---|---|---|---|---|---|---|
| Sol# | SnCl4 | F-127 | HCl | Ethanol | Sol# | Zn-acac | F-127 | NH4OH | Ethanol |
| Sn-1 | 1 | 0.008 | 2.21 | 21.7 | Zn-1 | 1 | 0.02 | 18.8 | 33.4 |
| Sn-2 | 1 | 0.04 | 7.7 | 39.6 | | | | | |

## 3. Results and Discussion

*3.1 Nanomaterial Refractive Index Exploration*

In **Figure 2A** the Ellipsometry determined refractive indices of $TiO_2$ prepared by precursor solutions Ti-1 to Ti-5, annealed at 400°C in air, along with refractive indices for a solution processed dense $TiO_2$, are shown. The refractive indices by CVD were obtained from the literature[34, 35] and displayed for comparison. **Figure 2B** shows the refractive indices of the prepared $SiO_2$ precursor solutions, with increasing refractive indices from low to high. The processed precursor solutions for $SnO_2$ and ZnO provided the refractive indices shown in **Figure 2C**. The polymer concentration has the effect demonstrated in **Figure 2D** on the refractive indices, reaching a minimum at a certain mole fraction. The precursor solutions T-1 through Ti-4 provide decreasing refractive index values, whereas the refractive indices of samples prepared by Ti-5 were increasing. The precursor solution Ti-4 provides an approximate location of the minimum refractive index that is achievable with the current method. The complete removal of the block-copolymer at temperatures ~350°C densify the nanomaterials with subsequent heating at higher temperatures. Therefore, an annealing study was performed to explore the effect of the annealing temperature on the refractive indices (**Figure 2C**). It is observed that samples annealed at 800°C had refractive indices that were ~7% greater. Therefore, not too significant of change was observed in the refractive indices due to densification by annealing at higher temperatures at which the block-copolymer support structure is no longer present. In **Figure 3** example prepared films annealed at 600°C are shown to demonstrate the high optical quality. In general, the scale of the nano-features are <10nm, therefore, there is a scale separation of 40λ which should result in minimal scattering for light at the visible and NIR frequencies.

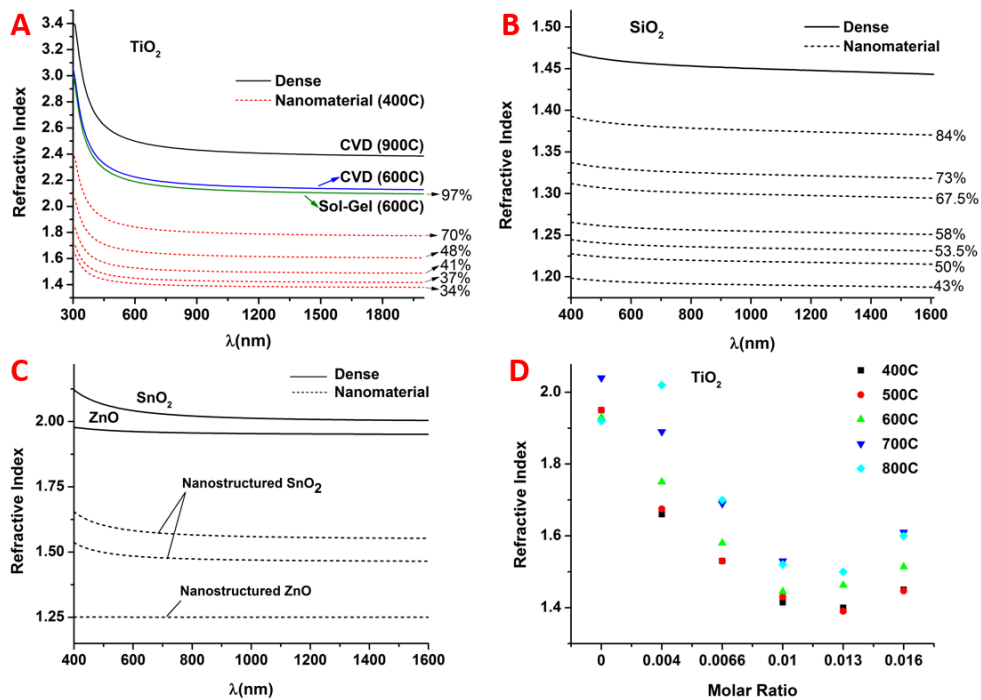

**Figure 2. A:** TiO$_2$ refractive indices provided by precursors Ti-1 through Ti-5 along with the refractive indices obtained for the dense TiO$_2$ precursor. In addition, refractive indices obtained from literature for TiO$_2$ deposited by CVD are included for comparison. The number percent displayed next to a line indicates the volume fraction of material with respect to air. **B:** SiO$_2$ refractive indices obtained from precursors Si-1 through Si-7. **C:** SnO$_2$ and ZnO refractive indices obtained by the developed precursor solutions. **D:** Examination of variations in the refractive indices as a function of annealing temperature and block-copolymer to Ti molar ratio.

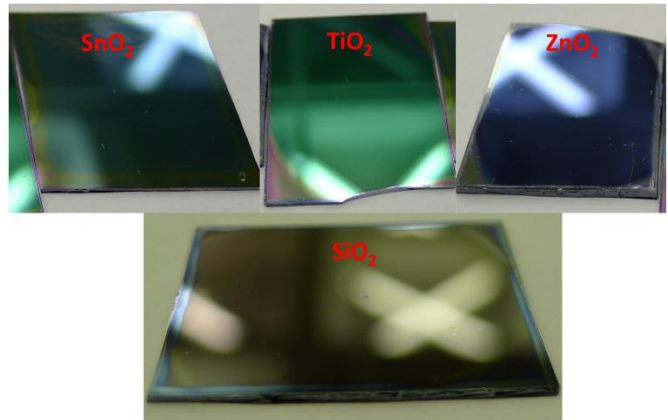

**Figure 3:** Images of the prepared nanomaterial oxides on silicon wafer showing high optical quality, obtained with precursors providing the lowest refractive indices.

## 3.2 Mesoporous Structure Visualization

To examine the structure formed from the various nanomaterials precursor solution, Ti-4, Sn-2, and the solution for ZnO were deposited on TEM grids with 9x9 windows of 0.1x0.1mm, with 50nm thick silicon nitride membranes. Bright field images were obtained with a JEOL JEM2100F TEM for the various nanostructures (**Figures 4 A-C**). A large degree of porosity is evident with an average grain size of <10nm for $TiO_2$ and $SnO_2$. On the other hand, the bright field TEM for ZnO shows an average grain size ~15-20nm. In **Figure 4D** an SEM image of a ZnO sample prepared on silicon wafer is shown for comparison. With TEM imaging, the oxide grains appear as the darker regions whereas in SEM imaging, they appear as lighter regions.

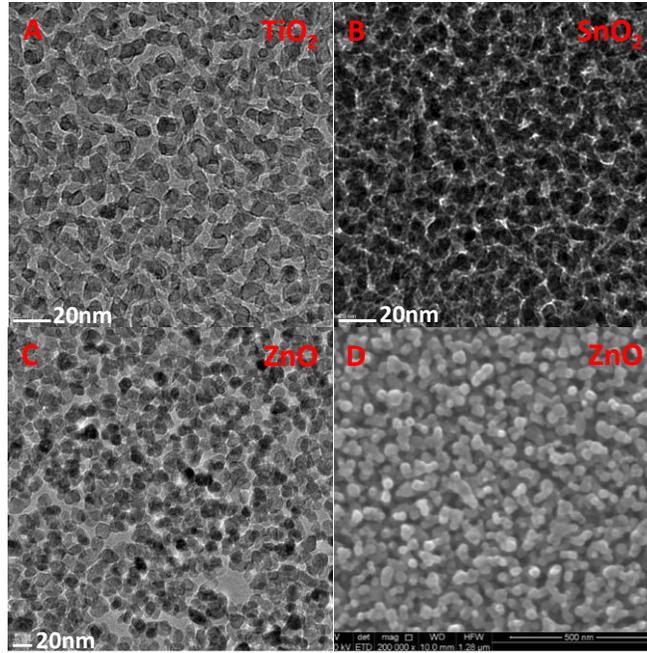

**Figure 4:** Bright field TEM images for **A:** $TiO_2$ **B:** $SnO_2$, and **C:** ZnO prepared on silicon nitride TEM grids **D:** SEM image or the surface of a ZnO sample prepared on silicon wafer. The images correspond to precursors yielding the lowest refractive indices.

## 3.3 Nanomaterial Anti-Reflection Coating

The refractive indices and film thicknesses provided by the transfer matrix method coupled with the optimization techniques used to obtain global and omnidirectional reflectivity minimums, were manufactured on silicon. Two coatings of the solution containing titanium isopropoxide, monoethanolamine, and 2-methoxyethnaol with molar ratio of 1:2:30 were spin cast at 3000RPM to obtain the dense $TiO_2$ layer. After each coating, the samples were placed on a hotplate set to 300°C, for 10 minutes, followed by the deposition of the $SiO_2$ precursor providing a refractive index near 1.22. The as prepared samples were then annealed at 600°C, as described before. With Ellipsometry the thickness of the dense $TiO_2$ layer in contact with the silicon substrate was measured to be 59nm and the thickness of the $SiO_2$ layer was measured to be 107nm. Practically, no measurable difference was noted in the reflectance of the two-layer coating measured at angles of incidence of 17° and 42°. From the measured reflectance spectra, the average reflectivity was estimated to be ~3%. This compares reasonably well with the design value of 2.82%, given that there is some degree of uncertainty in estimating the refractive indices of porous nanomaterials by Ellipsometry due to modelling and the slight differences in the obtained film thicknesses.

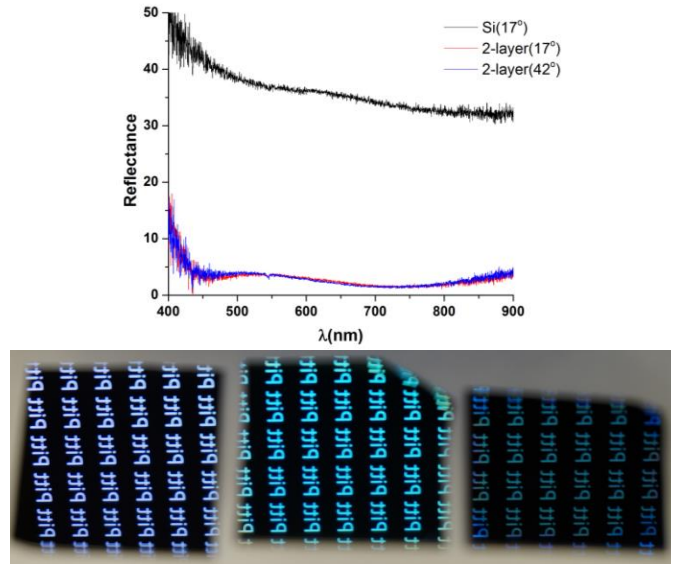

**Figure 5: Top:** Measured reflectance (%) of crystalline silicon and a two-layer anti-reflection coating. **Bottom:** Photograph of an image projected onto crystalline silicon, a one-layer $TiO_2$ coating, and the manufactured two-layer coating, showing reductions in the reflected visible light.

## 4. Conclusion

A method to engineer the refractive indices of functional materials by 3D nanostructuring in the deep sub-wavelength regime is demonstrated. The method is based on block-copolymer templating coupled with a low-cost wet processing approach to provide functional optical materials with tailorable refractive indices in the range of 1.17 to 2.2. The capability to tailor the refractive indices has potential high impact applications in a variety of fields.

In the explored application of anti-reflection coating, it is shown that with a two-layer coating light reflections at a silicon air interface were reduced from ~38% to 3%. Further reductions in the reflected light are possible by surface structuring and with a combination of increasing the highest achievable refractive index of $TiO_2$ (currently 2.2) and by reducing the lowest achievable refractive index of $SiO_2$ (currently 1.17). These improvements would allow for the coating of additional layers, providing further reductions in the reflected light between silicon air interfaces. When combined with techniques such as spray deposition, the realization of continuous refractive index gradients could potentially be realized by altering the concentration of the block-copolymer in time.

As recently demonstrated, the transformation optics design methodology can be used to design various conceptual optical components, such as efficient light distribution, bending, and coupling devices for light wave/silicon photonic circuits[14]. Transformation optics is a tensor based design technique and, therefore, it generally provides designs requiring three-dimensionally distributed strong refractive index gradients. It is needless to say that such designs are very difficult, if not impossible, to manufacture with existing technologies. It may be possible to realize many transformation optics designs by combining solution based refractive index engineering with such techniques as dip pen lithography, or other compatible technologies. With such a combination, 3D refractive index gradients could be printed with a possible resolution down to 50nm[40].

**Acknowledgements**

This work was supported by the National Science Foundation (CMMI-1054652, and CMMI-1300273) and the Department of Energy (DE-FE0003859)).

This report was prepared as an account of work sponsored by an agency of the United States Government. Neither the United States Government nor any agency thereof, nor any of their employees, makes any warranty, express or implied, or assumes any legal liability or responsibility for the accuracy, completeness, or usefulness of any information, apparatus, product, or process disclosed, or represents that its use would not infringe privately owned rights. Reference herein to any specific commercial product, process, or service by trade name, trademark, manufacturer, or otherwise does not necessarily constitute or imply its endorsement, recommendation, or favouring by the United States Government or any agency thereof. The views and opinions of authors expressed herein do not necessarily state or reflect those of the United States Government or any agency thereof.